\documentclass[reprint,amsmath,amssymb,aps,pra,showkeys,showpacs]{revtex4-1}
\usepackage{graphicx}
\usepackage{color}

\newcommand{\ket}[1]{\mathinner{|{#1}\rangle}}
\newcommand{\bra}[1]{\mathinner{\langle{#1}|}}

\begin{document}
\title{Dynamic generation of GHZ states with coupled charge qubits}

\author{J. Nogueira}
\affiliation{Instituto de F\'{\i}sica, Universidade Federal de Uberl\^andia, Av. Jo\^ao Naves de \'{A}vila, 2121 - Santa M\^onica, Uberl\^andia - MG, 38408-100, Brazil}
\author{P. A. Oliveira}
\affiliation{Instituto de F\'{\i}sica, Universidade Federal de Uberl\^andia, Av. Jo\^ao Naves de \'{A}vila, 2121 - Santa M\^onica, Uberl\^andia - MG, 38408-100, Brazil}
\author{F. M. Souza}
\affiliation{Instituto de F\'{\i}sica, Universidade Federal de Uberl\^andia, Av. Jo\^ao Naves de \'{A}vila, 2121 - Santa M\^onica, Uberl\^andia - MG, 38408-100, Brazil}
\author{L. Sanz}
\affiliation{Instituto de F\'{\i}sica, Universidade Federal de Uberl\^andia, Av. Jo\^ao Naves de \'{A}vila, 2121 - Santa M\^onica, Uberl\^andia - MG, 38408-100, Brazil}
\email{lsanz@ufu.br}
\date{\today}

\begin{abstract}
In this paper, we present a proof-of-principle of the formation of pure maximally entangled states from the Greenberger-Horne-Zeilinger class, in the experimental context of charged quantum dots. Each qubit must be identified as a pair of quantum dots, sharing an excess electron, coupled by tunneling. The electron-electron interaction is accounted for and is responsible for the coupling between the qubits. The interplay between coherent tunneling events and many-body interaction gives rise to the formation of highly entangled states. We begin by treating the problem of encoding three-qubits in a system with three pairs of quantum dots, and the numerical analysis of the exact quantum dynamics to find the conditions for the generation of the GHZ states. An effective two-level model sheds light on the role of a high-order tunneling process behind the dynamics. The action of the main decoherence process, the charge dephasing, is quantified in the process.  We then evaluate the physical requirements for the dynamical generation of GHZ states in a $N$ qubit scenario, and its challenges.

%\item[PACS numbers]
%\item[Keywords]
%GHZ states, entanglement, charge qubits, quantum dots
\end{abstract}
\maketitle

\section{Introduction}
\label{sec:intro}

Since the rise of the research on quantum information at the end of the XX century~\cite{Loss98,Burkard99}, semiconductor nanostructures have been pointed out as an interesting platform for the experimental implementation of quantum information processing. After twenty years, these at first theoretical expectations has been gradually fulfilled, with successful experimental realizations as the recent implementation of the fastest two-qubit gate using a 2D electronic gas in silicon~\cite{Zajac18,He2019}. Although electron spin $1/2$ states are the most common system to encode qubits~\cite{Yoneda18}, semiconductors offer other possibilities as the electronic state qubits defined in charged quantum dots~\cite{Hayashi03,Shinkai09}, the singlet-triplet qubit states of two-electrons in GaAs~\cite{Petta2005,Nichol2017}, and the exchange-only qubit with spin states~\cite{Laird10}, among others.

Quantum Dots (QDs) have been defined as artificial atoms, once the spatial confinement favors the formation of a discrete spectrum of electronic levels~\cite{Ashoori96}. From all the possibilities of encoding a qubit, as in the excitonic states~\cite{Scholl19,Borges16,Borges12} and the electronic spin~\cite{Russ18,Bugu20}, the interest on the physics of charged quantum dots has been increasing, once they are scalable systems where initialization and readout are possible through a process involving detection even of a single electron~\cite{Kiyama18,park2002}. In this physical system, the qubits are defined based on the property of electronic tunneling~\cite{Shinkai07,Shinkai09}, with the single-qubit operations being controlled by external gate voltages~\cite{Shinkai07,Shinkai09}. Moreover, the single-molecule electronics has been an outstanding new field of research, due to its future feasible implementations as the construction of a cheaper and faster single-electron transistor~\cite{xiang2016,Ratner05}.

In this work, we encode a qubit using a pair of charged quantum dots sharing one electron in excess, forming a quantum molecule. Inside the molecule, the single-electron tunneling and the electronic detuning guarantees the qubit encoding, while electrostatic interaction with the electron in an adjacent similar system couples two qubits. Recently, proposals of a generation of the W states, another tripartite entangled state, had been studied in the context of spin-qubits~\cite{Bugu20} and superconductor qubits~\cite{Stojanovic20}. Here, our main goal is to establish physical conditions for the generation of genuine multipartite maximally entangled states, belonging to the GHZ class~\cite{Greenberger89}, in front of the results for the case of two charge qubits~\cite{Souza19,souza2017,Oliveira15}. We are interested not only in explore, by numerical simulations, this dynamic, but also in the comprehension of the specific process behind the formation of the states and the feasibility of the proposal, in terms of the effects of the decoherence  process due to charge dephasing. As far as we know, this is the first theoretical demonstration that many-body interactions yields to highly entangled GHZ states in semiconductor charged quantum dots.

This paper is organized as follows. In section~\ref{sec:model_measures}, we present the model used to describe our system of interest, introducing the Hamiltonian operator for the physical setup and then showing the conditions for encoding three qubits. We also present some entanglement quantifiers that will be used for the characterization of the GHZ-class states. Section~\ref{sec:analysis} is devoted to the presentation of our main results: the numerical simulations for the generation of GHZ states together with the discussion about an effective two-level hamiltonian which illustrates how a three-order tunneling process explains the formation of the target states. The action of charge dephasing, the main mechanism of decoherence in our system of interest, is discussed in Sec.~\ref{sec:dephasing}. The scalability of our proposal for $N$ qubits encoded in quantum molecules is discussed in Sec.~\ref{sec:Nqubits}, and the Sec.~\ref{sec:summary} contains our final remarks.
\section{Model and entanglement quantifiers}
\label{sec:model_measures}
\begin{figure}[h]
    \includegraphics[scale=1]{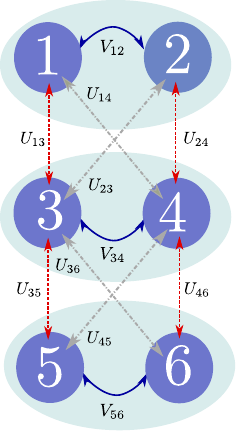}
    \caption{(Color online) Sketch of the physical system of three qubits: six quantum dots are coupled by pairs, each pair being a quantum molecule. Electronic coherent tunneling are permitted inside each molecule, although is forbidden between dots belonging to different molecules. The allowed tunneling couplings are shown with blue solid lines. Electrostatic coupling permits the interaction between electrons in different molecules, and are indicated by dot and dashed dot lines.}
    \label{fig:expsetup}
\end{figure}

In this section, we present the physical system considered, its model Hamiltonian, and a summary of the entanglement quantifiers. We begin with a model consisting of a three-qubit system, codified in charged quantum dots. A sketch of this particular physical system is shown in Fig.~\ref{fig:expsetup}. Each qubit is encoded in the electronic states of a double dot molecular structure~\cite{Hayashi03}. By attaching particle reservoirs to the molecules, the system can be initialized in some specific charge configuration, on demand. Additionally, Coulomb interaction between molecules is considered at least for first neighbors~\cite{Shinkai09,Fujisawa11}. This kind of multiple dots structure has been experimentally explored in the last two decades~\cite{Fujisawa98,Ka2020}.

Charged quantum dots are actually built at the intersection of two semiconductors of different band gaps (e.g. GaAs/AlGaAs), which sustains a two-dimensional electron gas (2DEG). The QDs are then delimited by spatial confinement obtained by the action of negative biased voltages on gate electrodes at the top of the 2DEG. Additionally, some extra gates couple the system with electronic reservoirs (not shown in the scheme): a source provides electrons to enter in the system and a drain withdraw charges, in a process that permits, for example, measurements of the current through each qubit. Extra electrodes attached at various points of the system are also used to attain a fine control of the physical parameters of tunneling, $V_{ij}$, and electronic energies, $E_i$, by simply varying gate voltages~\cite{Kouwenhoven91,Fujisawa98,Hayashi03,Shinkai09}.

We consider a closed system that is already initialized in one of the eight possible states for three qubits. We assume that the system is in the Coulomb blockade regime, where only a single electron per qubit is allowed. Our model Hamiltonian is written in second quantization as
\begin{equation}
\label{defH6qds}
\hat{H}_{6\mathrm{QDs}}=\sum_{i=1}^{6} E_{i}\hat{n}_{i}+\hat{V}+\hat{U},
\end{equation}
where
\begin{eqnarray}
\hat{V}&=&V_{12}\hat{d}_{1}^{\dagger} \hat{d}_{2}+V_{34}\hat{d}_{3}^{\dagger} \hat{d}_{4}  + V_{56}\hat{d}_{5}^{\dagger} \hat{d}_{6} +\mathrm{h.c.} \label{V}\\
\hat{U}&=&U_{13}\hat{n}_{1}\hat{n}_{3} + U_{14}\hat{n}_{1}\hat{n}_{4}+ U_{23}\hat{n}_{2}\hat{n}_{3}+U_{24}\hat{n}_{2}\hat{n}_{4}\nonumber\\
&&+U_{35}\hat{n}_{3}\hat{n}_{5} + U_{36}\hat{n}_{3}\hat{n}_{6} + U_{45}\hat{n}_{4}\hat{n}_{5} + U_{46}\hat{n}_{4}\hat{n}_{6},\label{U}
\end{eqnarray}
with $\hat{d}_{i}^{\dagger}$ ($\hat{d}_{i}$) being the fermionic creation (annihilation) operator, and  $\hat{n}_{i}= \hat{d}_{i}^{\dagger} \hat{d}_{i}$ is the number operator to the $i$th dot. The first term in Eq.~(\ref{defH6qds}) is the energy of the electronic level in each dot. The second, given by Eq. (\ref{V}), describes the intra molecule tunnel coupling, while the last term, Eq.(\ref{U}), accounts for the Coulomb interaction between molecules.

The Hilbert space $\mathcal{H}_{6\mathrm{QDs}}$ of the system is a $64$-dimensional space with elements being $\{\ket{1}^{\otimes 6},...,\ket{0}^{\otimes 6}\}$, and $\ket{1}$ ($\ket{0}$) standing for the dot being occupied (empty). Imposing the condition where a single electron per qubit is considered, the accessible Hilbert space reduces to only 8 states, with each molecule occupying one of the states $\ket{10}$ or $\ket{01}$. Thus, defining $\ket{10}=\ket{0}$ and $\ket{01}=\ket{1}$, the complete basis turns out to be $\{\ket{000}, \ket{001}, \ket{010}, \ket{011},\ket{100},\ket{101},\ket{110},\ket{111}\}$.

At this point, we introduce a new set of physical parameters in connection with the energies and couplings of Hamiltonian of Eq.(\ref{defH6qds}):
\begin{eqnarray}
\label{eq:param6qd3qb}
\varepsilon_1&=&{(E_1-E_2)}/{2},\;\varepsilon_2={(E_3-E_4)}/{2},\;\varepsilon_3={(E_5-E_6)}/{2},\nonumber\\
\Delta_1&=&-V_{12}=-V_{21},\;\Delta_2=-V_{34}=-V_{43},\nonumber\\
\Delta_3&=&-V_{56}=-V_{65},\\
J_{12}&=&U_{13}=U_{24}=- U_{14}=-U_{23},\nonumber \\
J_{23}&=&U_{35}=U_{46}=-U_{36}=-U_{45}.\nonumber
\end{eqnarray}
with this definitions and using only the reduced computational basis $\{\ket{0}^{\otimes 3},...\ket{1}^{\otimes 3}\}$, we can write the Hamiltonian in the following way
\begin{equation}
\label{ham3qbger}
\hat{H}_{3\mathrm{qb}}=\sum_{q=1}^{3} [\varepsilon_{q}\hat{\sigma}_{z}^{(q)}+\Delta_{q}\hat{\sigma}_{x}^{(q)}] + \sum_{q',q''} J_{q'q''}\hat{\sigma}^{(q')}_{z}\otimes \hat{\sigma}^{(q'')}_{z},
\end{equation}
where $\sigma_{z} = \left|0\rangle\langle 0\right|- \left|1\rangle\langle 1\right|$, and $\sigma_{x} =\left|0\rangle\langle 1\right|+ \left|1\rangle\langle 0\right|$. Here, the index $q=1,2,3$ denotes the qubits in the first two terms while $q'=1,2$ with $q''=q'+1$ are used in the third term. The first term in Eq.(\ref{ham3qbger}) depends on the detuning of the electronic energies for each qubit, the second takes into account the tunneling process inside each qubit, and the third term describes the Coulomb interaction between qubits. Notice that this Hamiltonian is an extension of the case for a two-qubit system in the context of charged quantum dots studied in earlier works~\cite{Oliveira15,souza2017,Souza19}.

As the number of qubits encoded in a quantum system increases, to determine the entanglement degree of quantum states becomes a challenge. To the well-known case of a two-qubit system, one can easily characterize the physical states as being separable or entangled. A similar analysis is not that simple, even for the case of three qubits~\cite{horodeckireview,dur00,sabin08}. Apart from the fully separable states, there are three classes of entangled states. We can say that two different entangled states belong to the same equivalence class if it is possible to find a set of Stochastic Local Operations and Classical Communication (SLOCC) that transform one state into another. Thus, concerning the classes of entangled states for three qubits, the first is the class of biseparable states, where one qubit remains separate while the remain two qubits show bipartite entanglement. The second and the third classes, the W and GHZ states, are two different families that show genuine tripartite entanglement~\cite{dur00}.

We are interested on studying the dynamical formation of states belonging to the GHZ class. For instance, we can write
\begin{equation}
\label{defGHZstate}
\ket{\Psi_{\mathrm{GHZ}}(\phi)}=\frac{1}{\sqrt{2}}\left(\ket{000} + e^{i \phi} \ket{111}\right),
\end{equation}
where $\phi$ is a relative phase. Notice that the application of the operator $I \otimes \sigma_x \otimes I$ on $\ket{\Psi_{\mathrm{GHZ}}}$ results in a new state of the form
\begin{equation}
\label{defFLIPstate}
\ket{\Psi_{\mathrm{FLIP}}(\phi)}=\frac{1}{\sqrt{2}}\left(\ket{010}+ e^{i \phi}\ket{101}\right),
\end{equation}
which belongs to the same GHZ class, once the operation is an invertible local operator (ILO). This means that the states in Eqs. (\ref{defGHZstate})-(\ref{defFLIPstate}) are SLOCC equivalent. In what follows, we will focus on the formation of both GHZ and FLIP states. This class of entangled states is interesting since it is well defined for any number of a multipartite qubit system ($N>2$). A generic GHZ representative state is of the form
\begin{equation}
\label{defGHZstateN}
\ket{\Psi_{\mathrm{GHZ}}^N(\phi)}=\frac{1}{\sqrt{2}}\left(\ket{0}^{\otimes N} + e^{i \phi} \ket{1}^{\otimes N}\right),
\end{equation}
and analogously the $\ket{\Psi_{\mathrm{FLIP}}^N(\phi)}$ is obtained by applying the ILO operation $\hat{I}\otimes\hat{\sigma}_x\otimes\hat{I}\otimes\hat{\sigma}_x...$ over $\ket{\Psi^{\mathrm{GHZ}}_N(\phi)}$.

In order to characterize the dynamical formation of GHZ states of three qubits, we calculate the $3-$tangle ($\tau_{3}$), a useful entanglement quantifier~\cite{coffman00} which is defined as
\begin{equation}
\label{tangle3}
\tau_3=\tau_{A(BC)}-\tau_{AB}-\tau_{AC},
\end{equation}
with A, B, and C representing the qubits, and
\begin{equation}
\tau_{A(BC)}=4\mathrm{Det}\left(\hat{\rho}_{A}\right),
\end{equation}
where $\hat{\rho}_{A}$ is the reduced density operator obtained by taking the partial trace with respect to both $B$ and $C$
qubits, i.e., $\hat{\rho}_{A} = \mathrm{Tr}_{B}\{\mathrm{Tr}_{C} \{\hat{\rho}(t)\}\} $. For the calculation of $\tau_3$ it is also necessary to calculate
\begin{equation}
 \tau_{AB}=\mathrm{Tr} (\hat{\rho}_{AB} \tilde{\hat{\rho}}_{AB}) - 2\lambda_{1} \lambda_{2},
\end{equation}
where $\hat{\rho}_{AB} = \mathrm{Tr}_{C} \{\hat{\rho}(t)\}$, $\tilde{\hat{\rho}}_{AB}=(\hat{\sigma}_y \otimes \hat{\sigma}_y)\hat{\rho}_{AB}^{*}(\hat{\sigma}_y \otimes \hat{\sigma}_y)$ being the spin-flip density operator, and $\hat{\rho}_{AB}^{*}$ is the complex conjugate of $\hat{\rho}_{AB}$. The values $\lambda_1$ and $\lambda_2$ are the only non-null square root eigenvalues of the operator $\hat{\rho}_{AB}\tilde{\hat{\rho}}_{AB}$. A similar definition holds for $\tau_{AC}$. As an auxiliary quantity, we define
\begin{equation}
\tau_2=\tau_{AB}+\tau_{AC},
\end{equation}
which quantifies the amount of bipartite entanglement in the three qubit system. This entanglement quantifier is used in our numerical calculations to establish if some pure state $\hat{\rho}(t)$ belongs to the GHZ class of entangled states, with a $3-$tangle value reaching $\tau_{3}\left[\hat{\rho}(t)\right] = 1$ while $\tau_2 = 0$, showing genuine multipartite entanglement.

A second useful quantity, used to establish the distance between two quantum states, is the fidelity, which is given by
\begin{eqnarray}
\label{defFidelity}
\mathcal{F}= \mathrm{Tr} \left[\hat{\rho}(t)\hat{\rho}_{\mathrm{target}}\right].
\end{eqnarray}

The fidelity reaches $1$ as the evolved density matrix operator, $\hat{\rho}(t)$, approaches to the target state $\hat{\rho}_{\mathrm{target}}$. We will use the fidelity to check the dynamics of the formation of GHZ states for the case of three qubits and the analysis of the scalability of our proposal.
\section{Dynamical generation of GHZ states}
\label{sec:analysis}

In this section we discuss the main results of our work, being the dynamical generation of a state of three qubits with genuine multipartite entanglement, belonging to the GHZ class. The dynamics of the closed system is obtained by solving the Von Neumann equation for the density matrix operator $\hat{\rho}(t)$ ($\hbar=1$),
\begin{equation}
\label{defVonNeumann}
\dot{\hat{\rho}}(t)=-i[\hat{H}_{3\mathrm{qb}},\hat{\rho}].
\end{equation}
Once the evolved density matrix $\hat{\rho}(t)$ is obtained, we calculate the population of states $\ket{000}$ and $\ket{111}$ via
\begin{eqnarray}
P_{e}(t)=\mathrm{Tr}\left[\hat{\rho}(t)\hat{\rho}_e\right],
\end{eqnarray}
where $\hat{\rho}_e=\left|e\rangle\langle e\right|$ with $e=000$ or $111$. We also calculate the entanglement measurements defined in Sec.~\ref{sec:model_measures}, and the fidelity, choosing some target state from the GHZ class, considering a specific value of relative phase $\phi$ in Eq.~(\ref{defGHZstate}).

In our studies, the Coulomb strength will be fixed at $J_{12} = J_{23} = J = 25\mu eV$~\cite{Fujisawa11}, and we set all physical parameters in Hamiltonian (\ref{ham3qbger}) in terms of $J$. In order to choose the value of $\Delta$, we run simulations looking for a combination of both, low times for formation of the GHZ state and high fidelity. In the range of $J/10\le\Delta\le J/2$, the numerical estimation of the time of formation of a GHZ state falls from $20$ ns to $0.1$ ns. At the same time, high values of $\Delta$ show a decrease of the fidelity of evolved state with a GHZ target state, from $0.99$ to $0.91$. Additionally, the analysis of the characteristics of the energy spectrum and eigenstates of the Hamiltonian, Eq.(\ref{ham3qbger}), shows that this choice of parameters favors the formation of two eigenstates with fidelities above $0.9$ with GHZ states $\ket{\Psi_{\mathrm{GHZ}}(\pi)}$ and $\ket{\Psi_{\mathrm{GHZ}}(0)}$ (see Appendix~\ref{app:eigenproblem} for details). This analysis of the spectrum and dynamics shows that the best conditions for the generation of a GHZ state occur when the electronic levels are resonant ($\varepsilon_1=\varepsilon_2=\varepsilon_3=0$), and the molecules have equal tunneling couplings being $\Delta_1=\Delta_2=\Delta_3=\Delta$.

\begin{figure}[t]
\includegraphics[scale=0.48]{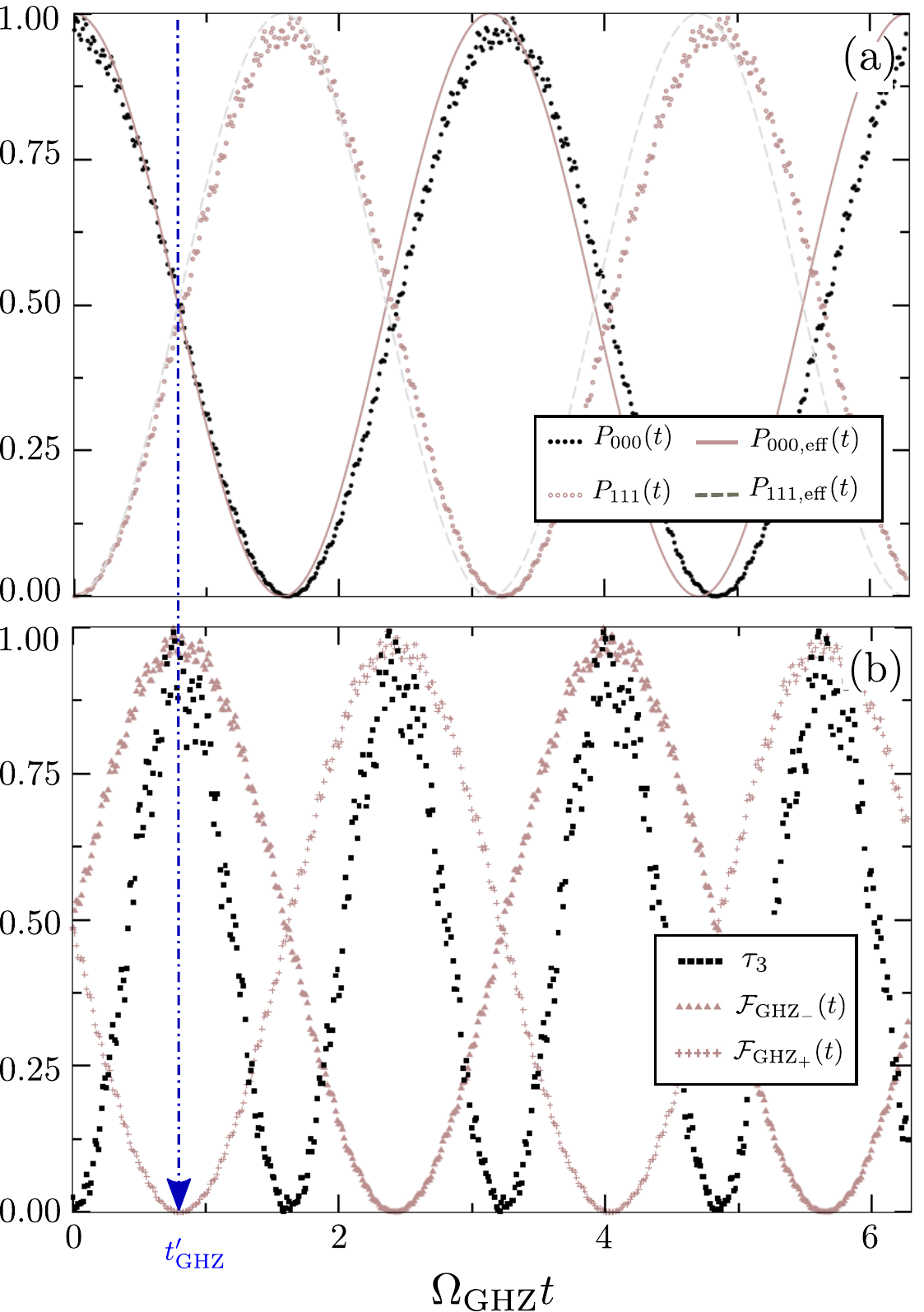}
\caption{(Color online) Quantum electronic dynamics in three quantum molecules, as a function of the dimensionless parameter $\Omega_{\mathrm{GHZ}} t$  considering the initial condition $\hat{\rho}(0)=\left|000\rangle\langle 000\right|$, and physical parameters given by $\varepsilon_q = 0$, $J=25 \mu eV$ and $\Delta_q = J/6$ in the Hamiltonian, Eq.(\ref{ham3qbger}). Panel (a) shows the populations $P_{000}(t)$ (black dots) and $P_{111}(t)$ (brown open circles), together with the evolution of the same quantities considering the effective two-level system as discussed in Sec.~\ref{sec:analysis}: $P_{000,\mathrm{eff}}(t)$ (brown solid line) and $P_{111,\mathrm{eff}}(t)$ (gray dashed line). Panel (b): Evolution of the entanglement quantifier $\tau_{3}$ (black squares), and the fidelities $\mathcal{F}_{\mathrm{GHZ}_-}$ (brown filled triangles), and $\mathcal{F}_{\mathrm{GHZ}_+}$ (brown crosses). The blue dash-dotted line shows the time of formation of the first GHZ state, corresponding to $t'_{\mathrm{GHZ}}=4.56$ ns as predicted by the effective two-level model.}
\label{fig:popDeltacase1}
\end{figure}

In order to keep both, short times of formation, important in order to face processes of decoherence, and high fidelity, we chose $\Delta=J/6$ for the illustration of our proof-of-principle of generation of GHZ states. Figure~\ref{fig:popDeltacase1}(a) shows both $P_{000}(t)$ and $P_{111}(t)$ against $\Omega_{\mathrm{GHZ}}t$, considering $\hat{\rho}(0) =\left|000\rangle\langle 000\right|$ as the initial condition. The choice of the coupling $\Omega_{\mathrm{GHZ}}$ to parametrize the temporal evolution will be made clear soon. Notice that when $P_{000}=P_{111}=0.5$ at $\Omega_{\mathrm{GHZ}}t'_{\mathrm{GHZ}}\approx \pi/4$, we obtain a highly entangled state, according to $\tau_3 = 1$, shown by the black squares in Fig.~\ref{fig:popDeltacase1}(b). For this particular time we also find $\tau_2 = 0$ (not shown here)~\footnote{The quantity $\tau_2$ is zero, although becomes non negligible for $\Delta>J/4$.}. To confirm the formation of GHZ states given by Eq.(\ref{defGHZstate}), we calculate the fidelities $\mathcal{F}_{\mathrm{GHZ}_-}$ (brown filled triangles), and $\mathcal{F}_{\mathrm{GHZ}_+}$ (brown crosses), which correspond to target states with $\phi=-\pi/2$ and $\phi=\pi/2$ in Eq.(\ref{defGHZstate}), respectively. Once, at time $t'_{\mathrm{GHZ}}$, the fidelity $\mathcal{F}_{\mathrm{GHZ}_-}$ becomes close to one, these results allow us to conclude that, at this particular time, the evolved state is given by
\begin{equation}\label{defGHZobtained}
\ket{\Psi(t'_{\mathrm{GHZ}})}\approx\ket{\Psi_{\mathrm{GHZ}}(-\pi/2)}=\frac{1}{\sqrt{2}}\left(\ket{000} -i \ket{111}\right).
\end{equation}
Also in Fig.~\ref{fig:popDeltacase1}(b), we find $\mathcal{F}_{\mathrm{GHZ}_+}$ close to one, for $\Omega_{\mathrm{GHZ}}t\approx 3\pi/4$, thus revealing the formation of $\ket{\Psi_{\mathrm{GHZ}}(\pi/2)}$. With the same choice of physical parameters but considering the  initial condition $\hat{\rho}(0)= \left|010\rangle\langle 010\right|$, we observe a similar behavior behind the formation of FLIP states (see Appendix~\ref{app:flipstates} for details). It is valid that, for the experimental value of $J=25$ $\mu$eV, the earliest time of generation of a GHZ is around $t'_{\mathrm{GHZ}}\approx 4.5$ ns.

Let us search for a two-level model which will provide an important insight on the generation of the GHZ states. We start rewriting the three-qubit Hamiltonian from Eq.~(\ref{ham3qbger}) as
\[\hat{H}_{3\mathrm{qb}}=\hat{H}_0+\hat{V},\]
where
\begin{equation}
\label{defH0}
 \hat{H}_{0}= \sum_{q=1}^{3} \varepsilon_{q}\hat{\sigma}_{z}^{(q)} + \sum_{q',q''} J_{q'q''}\hat{\sigma}^{(q')}_{z}\otimes \hat{\sigma}^{(q'')}_{z},
\end{equation}
is the diagonal term of the Hamiltonian. The energies $E^0_k$ of $\hat{H}_{0}$ are
\begin{eqnarray}
E^0_{000}&=&E^0_{111}=2 J, \label{levels1}\\
E^0_{010}&=&E^0_{101}=-2 J, \label{levels2}\\
E^0_{001}&=&E^0_{011}=E^0_{100}=E^0_{110}=0,
\label{levels3}
\end{eqnarray}
obtained from Eqs.(\ref{energies_H0}) in Appendix~\ref{app:eigenproblem}, with $\varepsilon_q = 0$ for all values of $q$, and $J_{12} = J_{23} = J $. As discussed also in Appendix~\ref{app:eigenproblem}, at this condition there are three subspaces energetically separated by a gap $|2J|$: the subspace spanned by $\{\ket{000},\ket{111}\}$, the one given by $\{\ket{010},\ket{101}\}$ and finally the largest one with $\{\ket{001},\ket{011},\ket{100},\ket{110}\}$.

The non-diagonal term of $\hat{H}_{3\mathrm{qb}}$
\begin{equation}
\label{perturbation}
\hat{V}=\Delta \sum_{i=1}^{3} \hat{\sigma}_{x}^{i},
\end{equation}
describes the action of tunneling. This coupling removes energy degeneracies of $\hat{H}_0$ inside each subspace. For $\Delta<J/4$, numerical calculations show that the eigenstates from the original subspace with $E=2J$ become GHZ states with $\phi=0$ and $\phi=\pi$ in Eq.~(\ref{defGHZstate}) (see Appendix~\ref{app:eigenproblem} for details).

At this point, we consider the tunneling coupling as a perturbation, in order to find an analytical expression for the effective two-level coupling, following a procedure used in a recent work~\cite{Souza19}. As we are interested in the formation of the GHZ state given by Eq. (\ref{defGHZstate}), we can assume that the system is initialized in one of the states $\ket{000}$ or $\ket{111}$. Since no relaxation mechanisms are present (charge dephasing will be accounted for in the next section), we expect a temporal evolution from the initial state to a coherent superposition of both, $\ket{000}$ and $\ket{111}$ states. However, no direct coupling between these two states is present. In order to undergo a quantum evolution inside this subspace, the system needs to perform virtual transitions via the other states of the three-qubit basis. This virtual mechanism can be explained via perturbation theory.

The calculation involves separate the original basis of three-qubits in two parts, where $A$ is a two-dimensional subspace with elements $\ket{000}$ and $\ket{111}$ and $B$ contains the remaining six elements. The matrix representation of the Hamiltonian which describes the problem can be seen as
\begin{equation}
\label{defHVperturGHZ}
 \tilde{\hat{H}}  = \tilde{\hat{H}}_{0}+\tilde{\hat{V}}= \begin{pmatrix}
          \tilde{\hat{H}}^{AA} & \tilde{\hat{H}}^{AB} \\
          \tilde{\hat{H}}^{BA} & \tilde{\hat{H}}^{BB}
         \end{pmatrix}.
\end{equation}
Following the steps detailed in the Appendix~\ref{app:calcOmega}, we arrive to a two-level effective hamiltonian written as
\begin{equation}
\label{eq:Heffective}
 \hat{H}_{\mathrm{eff}}^{\mathrm{GHZ}}=\Omega_{\mathrm{GHZ}} \left|111\rangle\langle 000\right| + h.c.,
\end{equation}
where $\Omega_{\mathrm{GHZ}}$ corresponds to
\begin{eqnarray}
\label{defOmega5}
\Omega_{\mathrm{GHZ}}&=& \sum_{k=1}^{8} \sum_{u=1}^{8} \bra{111}\tilde{H}^{AB}\left|{k}\rangle\langle{k}\right| (E - \tilde{\hat{H}}_0^{BB})^{-1} \hat{V}^{BB}\nonumber \\
&&\times(E - \tilde{\hat{H}}_0^{BB})^{-1} \left|{u}\rangle\langle{u}\right| \tilde{\hat{H}}^{BA} \ket{000}.
\end{eqnarray}
Here the indexes $k$ and $u$ run over all the states in the computational basis, and $\hat{V}^{BB}$ is the $6\times 6$ matrix representation of the perturbation $\hat{V}$ in the subspace $B$. Notice that the sequence of the operators $\tilde{H}^{AB}$, $\hat{V}^{BB}$, and $\hat{H}^{BA}$ in the numerator of Eq.~(\ref{defOmega5}) indicate that third-order processes are behind the emergence of anticrossing of $\ket{000}$ and $\ket{111}$. Expanding the sum and substituting the matrix elements, the expression becomes
\begin{eqnarray}
\Omega_{\mathrm{GHZ}}&=&\Delta^3\left[\frac{1}{E_{A}-E_{011}}\left(\frac{1}{E_{A}-E_{001}}+\frac{1}{E_{A}-E_{010}}\right)\right.\nonumber\\
&&\left.+\frac{1}{E_{A}-E_{101}}\left(\frac{1}{E_{A}-E_{001}}+\frac{1}{E_{A}-E_{100}}\right)\right.\\
&&\left.+\frac{1}{E_{A}-E_{110}}\left(\frac{1}{E_{A}-E_{010}}+\frac{1}{E_{A}-E_{100}}\right)\right].\nonumber
\end{eqnarray}
The term $E_A$ in the equation can be approximated as the eigenenergy of the unperturbed Hamiltonian $\tilde{H}_0$ in the subspace $A$, which is $E_A = 2J$ (or $E_A = -2J$, when describing the dynamics inside the FLIP space). Finally, we arrive to the expression of the effective coupling (in units of energy)
\begin{equation}
\label{defeffectivefreq}
\Omega_{\mathrm{GHZ}}=\frac{\Delta^3}{J^2}.
\end{equation}
By using the Eq. (\ref{defeffectivefreq}) inside Eq. (\ref{eq:Heffective}) and calculating the quantum dynamics, we obtain the Rabi oscillations for the effective populations $P_{000,\mathrm{eff}}(t)$ (brown solid line) and $P_{111,\mathrm{eff}}(t)$ (gray dashed line), plotted in Fig.~\ref{fig:popDeltacase1}. Notice that the model describes well the results of the exact calculation at short times. A similar behavior can be observed for the dynamics considering an initial condition for the formation of FLIP states, as shown in Fig.~\ref{fig:appflip} in Appendix~\ref{app:flipstates}.

From our numerical calculations, we see that the quantum dynamics do not behave as periodic and show ripples, both signs of a more complex dynamics, which involves the whole set of eight states on the unitary dynamics. Still, at short times and small tunneling rates, the effective model becomes a useful tool for the estimation of the value of the time of formation of a high fidelity GHZ state, which is
\begin{equation}
\label{eq:timeGHZ}
t'_{\mathrm{GHZ}}=\frac{\pi}{4}\frac{J^2}{\Delta^3}.
\end{equation}
With this expression, we estimate $t'_{\mathrm{GHZ}}=4.56$ns for the physical conditions in Fig.~\ref{fig:popDeltacase1}. This time scale is shown in the figure with the blue dash-dotted line, coinciding with the formation of the state $\ket{\Psi_{\mathrm{GHZ}}(-\pi/2)}$.
Following the same procedure, but considering the two-level system with energies $E=-2J$ as $A$, we arrive to an equivalent expression but the case of formation of FLIP states. The final result gives $\Omega_{\mathrm{FLIP}}=\Omega_{\mathrm{GHZ}}$, thus explaining the similarities between the dynamics of the two cases, as observed by comparing Fig.~\ref{fig:popDeltacase1} and Fig.~\ref{fig:appflip} (Appendix~\ref{app:flipstates}).

\section{Dynamical behavior under a dephasing channel}
\label{sec:dephasing}

In this section we discuss the effects of charge dephasing, the main decoherence process in the context of charged quantum dots~\cite{Zhang18,nielsenbook, wildebook}. To quantify the effect of dephasing on the generation of a GHZ state, we numerically solve a Lindblad master equation~\cite{Fujisawa11} written as
\begin{eqnarray}
\label{defLindblad}
\dot{\hat{\rho}}(t)&=&-i[\hat{H}_{3\mathrm{qb}},\hat{\rho}] + \frac{1}{2}\sum_{k=1}^{8} \Gamma_{k}\left(2 \hat{L}_{k} \hat{\rho}\hat{L}_{k}^{\dagger}-\hat{L}_{k}\hat{L}_{k}^{\dagger} \hat{\rho}\right.\nonumber\\
&&\left. - \hat{\rho}\hat{L}_{k}\hat{L}_{k}^{\dagger}\right),
\end{eqnarray}
where the $\Gamma_{k}$ are the rates associated with the dephasing channel in energy units, $\hat{L}_{k}$ are the jump operators, and the time scale of the dephasing process is given by $T_{\mathrm{deph}}=1/\gamma_{k}=h/\Gamma_{k}$, with $\gamma_{k}$ in Hertz. The physical agent behind this process is the background charge fluctuation. To model the effect of dephasing, we consider the operators
\begin{equation}
\hat{L}_k  =  \left|{k}\rangle\langle{k}\right|,
\end{equation}
in Eq.~(\ref{defLindblad}). Here $\ket{k}$ is one of the elements of the computational basis where $\hat{L}_1 =\left|{000}\rangle\langle{000}\right|$, $\hat{L}_2 = \left|{001}\rangle\langle{001}\right|$, and so on. We run our numerical simulation in order to solve the master equation (\ref{defLindblad}) considering $\Gamma_k=\Gamma_{\mathrm{deph}}$ for all $k$.

\begin{figure}[h]
    \includegraphics[scale=0.75]{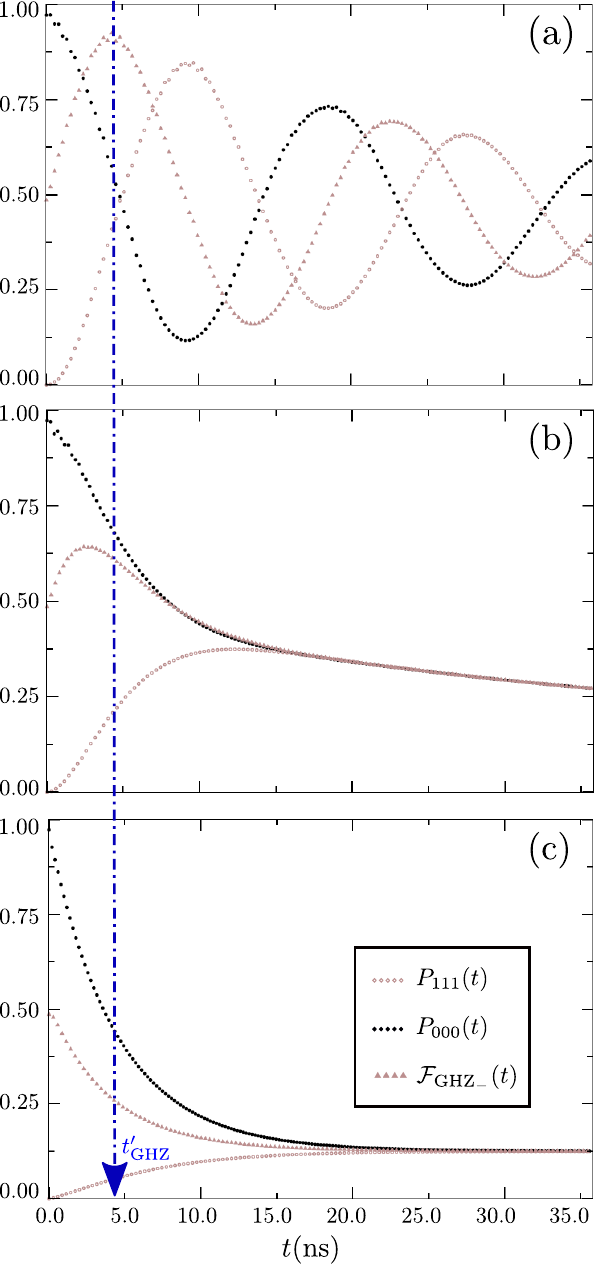}
    \caption{(Color online) Dynamics of populations $P_{000}(t)$ (black dots) and $P_{111}(t)$ (brown open circles), as well as the fidelity $\mathcal{F}_{\mathrm{GHZ}_-}$ (brown triangles) considering the action of the dephasing for $\varepsilon_q = 0$, $J=25 \mu eV$, $\Delta = J/6$, and the initial condition $\hat{\rho}(0)=\left|{000}\rangle\langle{000}\right|$ with (a) $\gamma_{\mathrm{deph}}=10^{-2}$ GHz; (b)$\gamma_{\mathrm{deph}}=10^{-1}$ GHz, and (c)$\gamma_{\mathrm{deph}}=1$ GHz. The blue dash-dotted line shows the time of formation of the first GHZ state, corresponding to $t'_{\mathrm{GHZ}}\approx 4.46 $ ns.}
    \label{fig:dephasing}
\end{figure}

In Fig.\ref{fig:dephasing}, we plot our results for the dynamics of populations $P_{000}(t)$ (black dots) and $P_{111}(t)$ (brown open circles), and the fidelity $\mathcal{F}_{\mathrm{GHZ}_-}$ (brown triangles), considering three different values of dephasing parameters: $\gamma_{\mathrm{deph}}=10^{-2}$ GHz in panel (a), $\gamma_{\mathrm{deph}}=10^{-1}$ GHz in panel (b), and $\gamma_{\mathrm{deph}}=1$ GHz in panel (c). The physical parameters are the same than in Fig.~\ref{fig:popDeltacase1}. We focus on the formation of the GHZ state $\ket{\Psi_{\mathrm{GHZ}}(-\pi/2)}$ once this state is generated earlier in the dynamics.
The results show the high sensibility of the three qubit dynamics over this process. Even for small rates of dephasing, Fig.~\ref{fig:dephasing}(a), the decreasing amplitude of oscillations is a clear sign that the process promotes population to others accessible states in a incoherent evolution. At $t'_{\mathrm{GHZ}}$, we still have the maximum for the fidelity with value $\mathcal{F}_{\mathrm{GHZ}_-}\approx 0.85$ for the target state $\ket{\Psi_{\mathrm{GHZ}}(-\pi/2)}$, although the populations at this time show a higher value for $P_{000}$ (around 0.62), which means the dephasing process affects strongly the third-order tunneling processes behind the effective two-level dynamics for the formation of the GHZ-class states. Increasing the value of the dephasing rate to $\gamma_{\mathrm{deph}}=10^{-1}$ GHz, shown in Fig.~\ref{fig:dephasing}(b), the oscillatory behavior is lost and the fidelity shows a maximum with low value ($\approx 0.55$). Finally, for $\gamma_{\mathrm{deph}}=1$ GHz, shown in Fig.~\ref{fig:dephasing}(c), the population $P_{000}$ and fidelity decay fast, with $P_{111}$ increasing, until $t\approx 12$ ns, when the density operator becomes a statistical mixture. In fact, for the three dephasing rates considered, it is true that $\lim_{t \to \infty}\hat{\rho}(t) = I/8$. This behavior shows that the feasibility of the generation of a GHZ class state in the context of charged qubits is an open experimental challenge, once it is important to guarantee a setup with low values of dephasing rates. Even though we find that the formation of GHZ states in the present system is quite sensitive to charge dephasing, if low dephasing rates are experimentally attained, one can find large fidelity values for both GHZ and FLIP states.

\section{Exploring the generation of GHZ states for $N$ qubits}
\label{sec:Nqubits}
After our careful analysis of the case with three qubits, we proceed to explore an extension to a general case of $N$ qubits, for which the dynamics is governed by the hamiltonian
\begin{equation}
\label{eq:HNqb}
\hat{H}_{N\mathrm{qb}}=\sum_{q=1}^{N} [\varepsilon_{q}\hat{\sigma}_{z}^{(q)}+\Delta_{q}\hat{\sigma}_{x}^{(q)}] + \sum_{q',q''} J_{q'q''}\hat{\sigma}^{(q')}_{z}\otimes \hat{\sigma}^{(q'')}_{z},
\end{equation}
where the physical parameters have the same interpretation: $2\varepsilon_q$ are the energetic detuning of the two electronic levels, and $\Delta_q$ accounts for the tunneling coupling inside a quantum molecule, while $J_{q'q''}$ describes the electrostatic interaction between electrons from different qubits.

\begin{figure}[h]
    \includegraphics[scale=0.42]{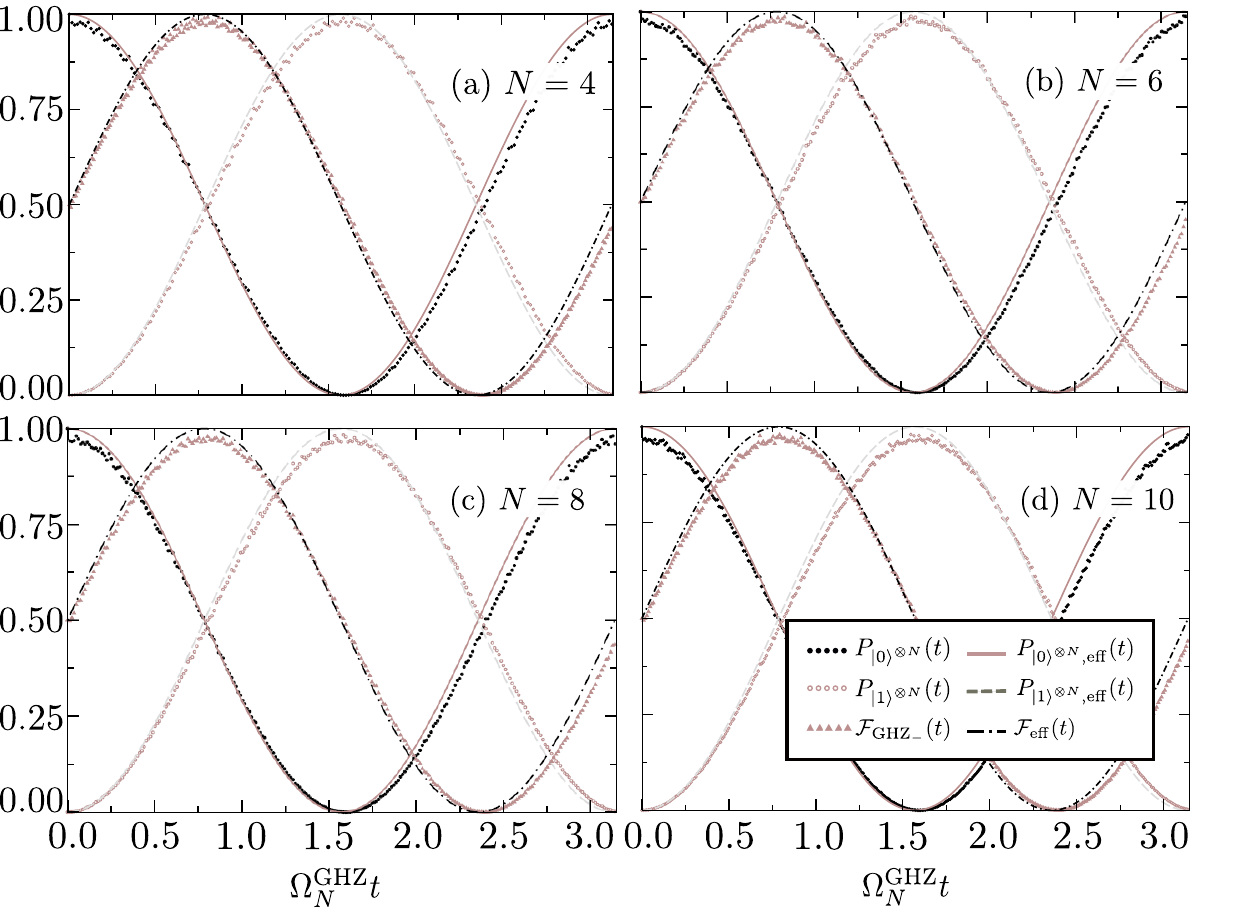}
    \caption{(Color online) Dynamics as function of populations $P_{\ket{0}^{\otimes N}}(t)$ (black dots) and $P_{\ket{1}^{\otimes N}}(t)$ (brown open circles), and the fidelity $\mathcal{F}_{\mathrm{GHZ}^N_{-}}$ (brown triangles) calculated with the target state $\ket{\Psi^{N}_{\mathrm{GHZ}}(-\pi/2)}$. We also plot the corresponding quantities calculated considering the effective two-level model: $P_{\ket{0}^{\otimes N},\mathrm{eff}}(t)$ (brown solid line), $P_{\ket{1}^{\otimes N},\mathrm{eff}}(t)$ (black dashed line), and the fidelity $\mathcal{F}_{\mathrm{GHZ}^N_{-,\mathrm{eff}}}$ (black dash-dotted line), also calculated with the target state $\ket{\Psi^{N}_{\mathrm{GHZ}}(-\pi/2)}$. The illustrated cases correspond to (a) $N=4$, (b)$N=6$, (c) $N=8$, and (d) $N=10$. Physical parameters are $\varepsilon_q = 0$, $J=25\mu$eV, $\Delta = J/8$, and the initial condition $\hat{\rho}(0)=\left|{0^{\otimes N}}\rangle\langle{0^{\otimes N}}\right|$.}
    \label{fig:dynamicsN}
\end{figure}

Calculations for the fidelity of the highest eigenstates of the case $N=4$ show that these states resemble the entangled states $\ket{\Psi_{\mathrm{GHZ}}^4(0)}$ (highest energetic, $16$th, eigenstate) and $\ket{\Psi_{\mathrm{GHZ}}^4(\pi)}$ ($15$th state), as shown in the Fig.~\ref{fig:fidspectrumN}(a). Careful analytical calculations of the applicability of the extension for $N=4$ of the two-level effective model result on the expression $\Omega^{\mathrm{GHZ}}_{4}=\frac{\Delta^4}{J^3}$~\footnote{The calculation of the value of the frequency of the effective two-level dynamics is done following a similar procedure as described on Appendix~\ref{app:calcOmega}, but considering the elements of subspace $A$ being $\left\{\ket{0}^{\otimes 4},\ket{1}^{\otimes 4}\right\}$.}, considering the same physical conditions for the detunings $\varepsilon_q=0$ and Coulomb parameters $J_{i,i+1}=J=25\mu$eV. The exact calculation of the dynamics for $N=4$ is shown in Fig.~\ref{fig:dynamicsN}(a), where we plot numerical results for the populations $P_{\ket{0}^{\otimes 4}}(t)$ (black dots) and $P_{\ket{1}^{\otimes 4}}(t)$ (brown open circles), and the fidelity $\mathcal{F}_{\mathrm{GHZ}^4_{-}}$ (brown triangles) calculated with the target state $\ket{\Psi^{4}_{\mathrm{GHZ}}(-\pi/2)}$. This dynamics is compared with those resulting from the generalized effective two-level model for $N=4$: $P_{\ket{0}^{\otimes N},\mathrm{eff}}(t)$ (brown solid line), $P_{\ket{1}^{\otimes N},\mathrm{eff}}(t)$ (black dashed line), and the fidelity calculated considering the two-level dynamics, $\mathcal{F}_{\mathrm{GHZ}^4_{-,\mathrm{eff}}}$ (black dash-dotted line). The results highlights that the state $\ket{\Psi_{\mathrm{GHZ}}^4(-\pi/2)}$ is generate at time \[t^{\mathrm{GHZ}}_4=\frac{\pi{J^{3}}}{4\Delta^4},\]
with $\mathcal{F}_{\mathrm{GHZ}^4_{-}}\sim 1$.

Based on this promising scenario, we propose an equation which follows the same structure of those obtained for $N=3$ and $N=4$ given by
\begin{equation}
t^{\mathrm{GHZ}}_N\sim\frac{\pi}{4\Omega^{\mathrm{GHZ}}_{N}}=\frac{\pi{J^{N-1}}}{4\Delta^N}.
\label{eq:timeGHZN}
\end{equation}
At this time, we expect the dynamical generation of a GHZ state of $N$ qubits as defined in Eq.~(\ref{defGHZstateN}) with $\phi=-\pi/2$. We check this proposition by comparing the exact dynamics, calculated numerically, and effective dynamics for values from $N=4$ to $N=10$. Without loss of generality, we choose the values $N=6,8$ and $N=10$ to illustrate the dynamics in Fig.~\ref{fig:dynamicsN}(b), Fig.~\ref{fig:dynamicsN}(c), and Fig.~\ref{fig:dynamicsN}(d) respectively. In all the cases presented, the fidelity remains above 0.98, thus indicating the robustness of the formation of the GHZ state for increasing number of qubits. The decreasing of fidelity can be understood by observing that the eigenstates of the two-level system with the highest energy show a decreasing fidelity with the highly entangled states $\ket{\Psi_{\mathrm{GHZ}}^N(0)}$ (for the $2^N$th state) and $\ket{\Psi_{\mathrm{GHZ}}^N(\pi)}$ (for the $(2^N-1)$th state), as shown in Fig.~\ref{fig:fidspectrumN} in Appendix~\ref{app:eigenproblem}.

Still, the main limitation on the generation of states of the GHZ class is dephasing. Notice that the time of generation $t^{\mathrm{GHZ}}_N$ scales with $\frac{\pi{J^{N-1}}}{\Delta^N}$. Once high fidelities are obtained for $\Delta=J/f$ with $f>1$, we can rewrite a temporal scale of formation of the GHZ state for $N$ qubit as $t^{\mathrm{GHZ}}_N(f,J)\sim\frac{\pi}{4}f^{N}J^{-1}$, which would be generally high, even for small values of $f$ and high values of $J$. For instance, if we consider $f=6$ and $J=25\mu$eV, the time scale of formation of the state with $N=6$ qubits  $\ket{\Psi^{6}_{\mathrm{GHZ}}(-\pi/2)}$ is around $1\mu$s. One can say that, in terms of scalability, we face a conundrum: a small value of $f$ means a larger value of $\Delta$, which impacts the fidelity of the GHZ state, but the state is more robust against dephasing once the time scale is shorter. On the other hand, obtaining a state with a high degree of fidelity can be a hard task to fulfill because of the action of dephasing.

\section{Summary}
\label{sec:summary}
In this work, we discuss the generation of genuine multipartite states belonging to the GHZ class, in the context of semiconductors quantum dots. We first encode three qubits in three pairs of charged quantum dots, each pair defining a quantum molecule. In the Coulomb blockade regime, a single electron can be injected into each pair of quantum dots. This excess electron jumps back and forth between the dots, thus encoding a qubit. Electrostatic interaction between the quantum molecules guarantees the coupling between qubits.

We demonstrate that the unitary dynamics of this system can be manipulated to generate states of the GHZ class at short times, considering the resonance of the electronic energies and equal values of tunneling rates. Depending on the setup of the initial state, it is possible to create a GHZ or a FLIP state, where the time of formation can be controlled by the value of tunneling coupling. Although this time decreases as the tunneling rate increases, the dynamic shows that high values of $\Delta$ are not convenient for the formation of the entangled state, once the dynamic starts to populate other electronic states. Our analytical and numerical work considering three qubits permits us to understand the origins of an effective two-level dynamics based on a third-order tunneling process, with the earlier time of formation of a GHZ-class being $\approx\frac{\pi J^2}{4\Delta^3}$. We simulate the action of charge dephasing, experimentally pointed out as the main mechanism of decoherence in charged quantum molecules. Our analysis of a situation considering $N$ qubits shows a promising scenario for the scalability of the dynamical generation of GHZ states, although increasing the number of qubits is a challenge, once the numerical calculation reveals that the time scale for the formation of the GHZ state also increases.

\section{Acknowledgments}
L.S. thanks Augusto M. Alcalde for the assistance with numerical calculations. This work was supported by CAPES, and the Brazilian National Institute of Science and Technology of Quantum Information (INCT-IQ), grant 465469/2014-0/CNPq.

\appendix
\section{Characteristics of the energy spectrum and the eigenstates of charged quantum dots Hamiltonian encoding qubits}
\label{app:eigenproblem}
In this appendix, we discuss considerations about the spectrum and eigenstates of the three-qubit system, described by Hamiltonian in Eq.~(\ref{ham3qbger}), and its extension considering $N$ qubits, Eq.~(\ref{eq:HNqb}).

Let us begin by writing the expressions for the diagonal terms of the first Hamiltonian, which are the terms of hamiltonian $\hat{H}_0$ in Eq.(\ref{defH0}):
\begin{eqnarray}
\label{energies_H0}
E^0_{000}&=&\varepsilon_1 + \varepsilon_2 + \varepsilon_3 + J_{12} + J_{23},\\
E^0_{001}&=&\varepsilon_1 + \varepsilon_2 - \varepsilon_3 + J_{12} - J_{23},\nonumber\\
E^0_{010}&=&\varepsilon_1 - \varepsilon_2 + \varepsilon_3 - J_{12} - J_{23},\nonumber\\
E^0_{011}&=&\varepsilon_1 - \varepsilon_2 - \varepsilon_3 - J_{12} + J_{23},\nonumber\\
E^0_{100}&=&- \varepsilon_1 + \varepsilon_2 + \varepsilon_3 - J_{12} + J_{23},\nonumber\\
E^0_{101}&=&-\varepsilon_1 + \varepsilon_2 - \varepsilon_3 - J_{12} - J_{23},\nonumber\\
E^0_{110}&=&-\varepsilon_1 - \varepsilon_2 + \varepsilon_3 + J_{12} - J_{23},\nonumber\\
E^0_{111}&=&-\varepsilon_1 - \varepsilon_2 - \varepsilon_3 + J_{12} + J_{23}.\nonumber
\end{eqnarray}
If we impose the condition of equal value for detunings, i.e. $\varepsilon_q=0$ above ($q=1,2,3$, being the number of qubit), and if the Coulomb coupling is given by $J_{12}=J_{23}=J$, we obtain only three different energy values given by $E=-2J,0,2J$. That means, under this particular choice of physical parameters, the spectrum of $\hat{H}_0$ shows energy degeneracy that permit the definition of three different subspaces:
\begin{itemize}
\item $\mathcal{H}_{E=2J}$ with degeneracy $2$, and eigenstates are given by $\left\{\ket{000},\ket{111}\right\}$;
\item $\mathcal{H}_{E=0}$ with degeneracy 4, and elements are given by $\left\{\ket{001},\ket{011},\ket{100},\ket{110}\right\}$;
\item $\mathcal{H}_{E=-2J}$ with degeneracy $2$ and eigenstates being $\left\{\ket{010},\ket{101}\right\}$.
\end{itemize}

\begin{figure}[h]
   \includegraphics[scale=0.5]{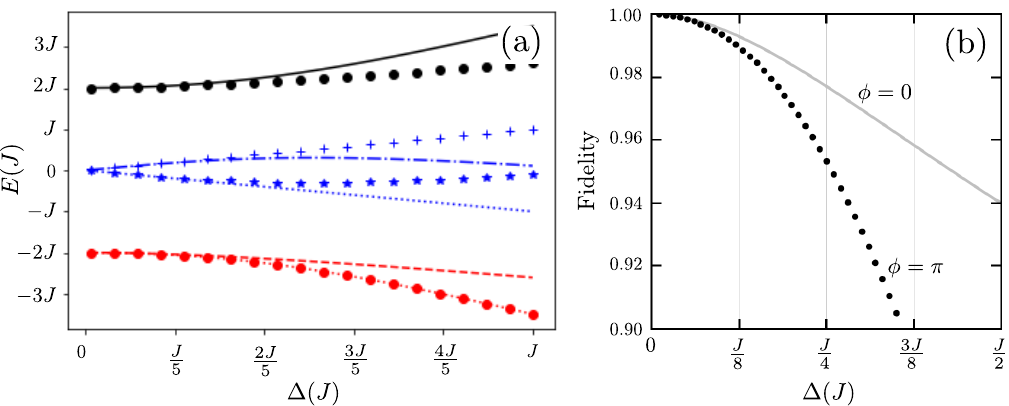}
    \caption{(Color online) Panel (a): the energy spectrum of the three-qubit Hamiltonian, Eq.~(\ref{ham3qbger}), against the tunneling coupling $\Delta$ considering $J=25\mu$eV, $\varepsilon_q=0$, and equal tunneling rates for the three quantum molecules. The values of energies and $\Delta$ are expressed in terms of $J$. The plot includes the $1$th eigenstate (red circles and dotted line), and the $2$nd (red dashed line), originally from the $\mathcal{H}_{E=-2J}$ subspace; the $3$rd (blue dotted line), $4$th (blue stars), $5$th (blue dot-dashed line), and $6$th (blue crosses) eigenstates, originally from the $\mathcal{H}_{E=0}$ subspace; and finally the $7$th (black circles) and $8$th (gray solid line) eigenstates, associated with the $\mathcal{H}_{E=2J}$ subspace. Panel (b): calculation of the fidelity of the states in subspace $\mathcal{H}_{E=2J}$ as function of $\Delta$, considering $\ket{\Psi_{\mathrm{GHZ}}(\pi)}$ as the target state for the $7$th state and $\ket{\Psi_{\mathrm{GHZ}}(0)}$ as the target state for the $8$th state. Line and symbols are the same used for the corresponding eigenvalues in panel (a).}
\label{fig:spectrum}
\end{figure}

We introduce the action of the tunneling, which corresponds to include the nondiagonal part of the three-qubit hamiltonian described by operator $\hat{V}$, Eq.(\ref{defHVperturGHZ}). In Fig.~\ref{fig:spectrum}(a) we plot the energy spectrum for $J=25\mu$eV against $\Delta$, considering $\Delta=\left.\left(0,J\right.\right]$. The energies are written in units of $J$ for the sake of clarity. The action of the tunneling is behind the emergence of anticrossings with the subsequent removal of degeneracy inside each subspace, which becomes important for values above $\Delta=J/2$.

At this point, we search for highly entangled eigenstates, as in previous works~\cite{Souza19,Oliveira15}, connected with anticrossings observed on the energy spectrum. For the subspace $\mathcal{H}_{E=2J}$, we seek for eigenstates given by
\begin{eqnarray}
\ket{\Psi_{\mathrm{GHZ}}(0)}&=&\frac{1}{\sqrt{2}}\left(\ket{000}+\ket{111}\right),\\
\ket{\Psi_{\mathrm{GHZ}}(\pi)}&=&\frac{1}{\sqrt{2}}\left(\ket{000}-\ket{111}\right).\nonumber
\end{eqnarray}
We proceed to calculate the fidelity of the eigenstates which emerges from subspace $\mathcal{H}_{E=2J}$ considering values for tunneling coupling given by $\Delta<J/2$. The results are shown in Fig.\ref{fig:spectrum}(b), where the high value for the fidelity, above $0.9$, corroborates that the eigenstates with the highest energies are approximately $\ket{\Psi_{\mathrm{GHZ}}(0)}$ and $\ket{\Psi_{\mathrm{GHZ}}(\pi)}$, although the fidelity decreases from $0.98$ to $\approx 0.90$ as $\Delta$ increases. A similar result (not shown here) for the fidelities of the ground and first excited states, the $1$st and $2$nd eigenstates, using as target states $\ket{\Psi_{\mathrm{FLIP}}(\pi)}$ and $\ket{\Psi_{\mathrm{FLIP}}(0)}$ respectively. This results give us confidence to search for an effective two-level model to describe the dynamics behind the formation of the GHZ states, considering the action of tunneling as a perturbation.

Concerning the potential scalability of the physical system, it is straightforward to find that the eigenstates of a hamiltonian of Eq.~(\ref{eq:HNqb}), considering no detuning ($\varepsilon_q=0$) and no tunneling ($\Delta_q=0$), now are organized in $N$ subspaces. Each subspace is associated with a value of energy given by $E_N=-(N-1)J,-(N-3)J,...,(N-3)J,(N-1)J$. We proceed to explore only the highest states of the Hamiltonian given by Eq.~(\ref{eq:HNqb}), considering situations with $N>3$. We now focus only on exploring the fidelity of the highest eigenstates considering as target states $\ket{\Psi_{\mathrm{GHZ}}^N(0)}$ (for the $2^N$th eigenstate) and $\ket{\Psi_{\mathrm{GHZ}}^N(\pi)}$ (for the $(2^N-1)$th eigenstate) in order to establish if there are similar anticrossings of highly entangled states as those found for $N=3$.

\begin{figure}[h]
   \includegraphics[scale=0.6]{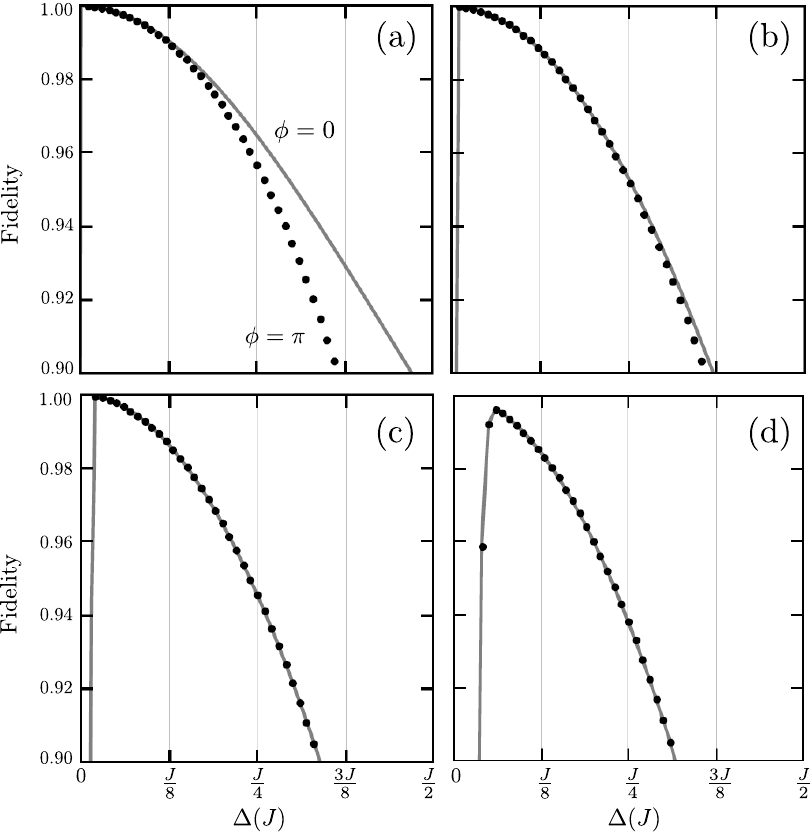}
    \caption{(Color online) Fidelity of the states in subspace $\mathcal{H}_{E=(N-1)J}$, as a function of tunneling parameter $\Delta$, considering $\ket{\Psi_{\mathrm{GHZ}}(\pi)}$ as the target state for the $(2^N-1)$th state (black circles) and $\ket{\Psi_{\mathrm{GHZ}}(0)}$ as the target state for the $2^N$th excited state (gray solid line) considering (a)$N=4$, (b) $N=6$, (c) $N=8$, and (d) $N=10$.}
\label{fig:fidspectrumN}
\end{figure}

In Fig.~\ref{fig:fidspectrumN}, we plot our results for the fidelities, calculated by exact diagonalization of Hamiltonian for $N$ qubits, Eq.(\ref{eq:HNqb}), considering $N=4$, $6$, $8$, and $N=10$ in panels (a), (b), (c), (d), respectively. Note that, for each $N$, both eigenstates lose fidelity with their correspondent target states as $\Delta$ increases. Basically, this is related to the fact that large $\Delta$ values tend to reduce the energy separation between these and the rest of the spectrum, becoming less similar to the GHZ states.

Also, the difference between the values of the fidelity values of $2^N$th and $(2^N-1)$th states with their target states starts to decrease as $N$ increases. In fact, for $N=10$, Fig.~\ref{fig:fidspectrumN}(d), both fidelities curves coincide. This behavior is explained by the fact that, as the dimension of the problem increases with $N$, small energetic differences between these states when $\Delta\neq 0$ has less impact on the entanglement degree of the eigenstates.

Another aspect that becomes evident from the Fig.~\ref{fig:fidspectrumN}(d) is that low values of tunneling coupling are not enough to guarantee entangled eigenstates. To favor the dynamical generation of GHZ state for $N$ qubits, a promising scenario is an experimental fine control of the tunneling coupling to keep enough high values of $\Delta$, which guarantees the formation of entangled eigenstates of the Hamiltonian of Eq.(\ref{defH0}) with high values of fidelities with GHZ states, but low enough to allow treating the tunneling as a perturbation.

\section{Generation of the FLIP states}
\label{app:flipstates}

\begin{figure}[h]
    \includegraphics[scale=0.45]{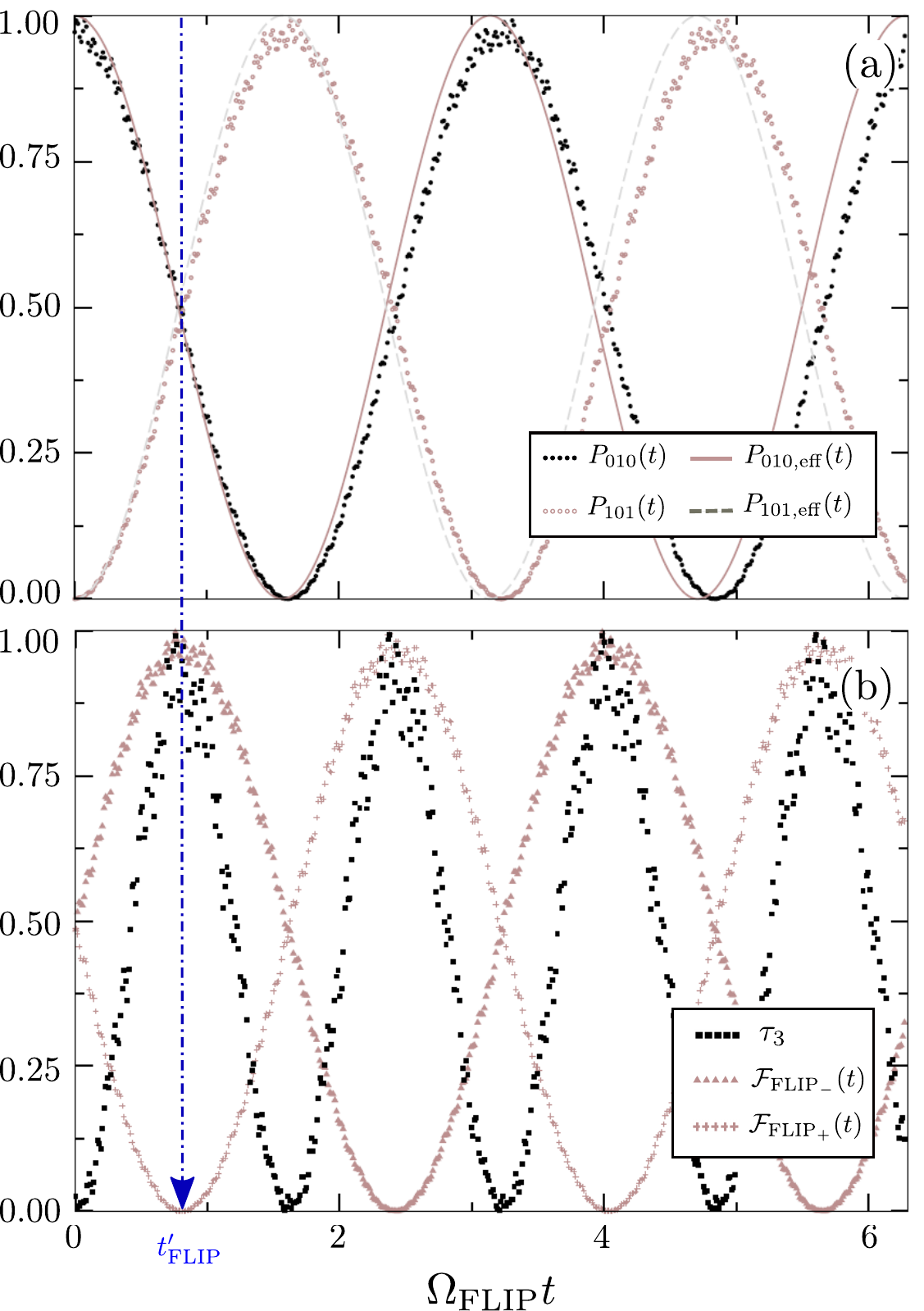}
    \caption{(Color online) Quantum dynamics against $\Omega_{\mathrm{FLIP}} t$ considering the initial condition  $\hat{\rho}(0)=\left|{010}\rangle\langle{010}\right|$ for $\varepsilon_q = 0$, $J=25 \mu eV$ and $\Delta = J/6$. Panel (a): Populations $P_{010}(t)$ (black dots) and $P_{101}(t)$ (brown open circles) and the dynamics of the effective two-level system for the population $P_{010,\mathrm{eff}}(t)$ (brown solid line) and $P_{101,\mathrm{eff}}(t)$ (gray dashed line). Panel (b): Evolution of  $\tau_{3}$ (black squares), and the fidelities $\mathcal{F}_{\mathrm{FLIP}_-}$ (brown filled triangles), and $\mathcal{F}_{\mathrm{FLIP}_+}$ (brown crosses). Fidelities are calculated using the Eq.(\ref{defFidelity}) for the target states $\ket{\Psi_{\mathrm{FLIP}}(-\pi/2)}$ and $\ket{\Psi_{\mathrm{FLIP}}(\pi/2)}$, respectively. The blue dash-dotted line shows the time of formation of the first GHZ state, corresponding to $t'_{\mathrm{FLIP}}\approx 4.56$ ns.}
    \label{fig:appflip}
\end{figure}

This appendix is devoted to present our results for the generation of the FLIP states, following the same procedure discussed in Sec.~\ref{sec:analysis} but considering the initial condition $\hat{\rho}(0) = \left|{010}\rangle\langle{010}\right|$, and the fidelity target states being $\hat{\rho}_{\mathrm{FLIP}(\pm\pi/2)}$. In Fig.~\ref{fig:appflip}, we plot our results for populations, fidelities, and $\tau_3$, panels below, considering  $\Delta=J/6$, the same value of Fig.~\ref{fig:popDeltacase1}.

By comparing this results with those discussed in the main text, we observe a similar behavior for populations,  entanglement degree, and fidelities with the main difference being that the dynamically accessed populations are now $P_{010}$ and $P_{101}$. The fidelity calculations and $\tau_{3}$ show that the state $\ket{\Psi_{\mathrm{FLIP}}(-\pi/2)}$ is generated at $t'_{\mathrm{FLIP}}=t'_{\mathrm{GHZ}}$. The continuous lines in panel (a) show the results of the effective two-level model, which coupling parameter is obtained following a similar process that those discussed on Sec.~\ref{sec:analysis} for the GHZ state.

\section{The nonlinear effect of the tunneling process and the two-level effective model.}
\label{app:calcOmega}
Once the analysis of the spectrum and eigenstates successfully points out the emergence of a well separate two-level system where the states $\ket{000}$ and $\ket{111}$ are coupled by tunneling considered as a perturbation, we are ready to understand the underlying processes behind the formation of the GHZ state. We first split $\mathcal{H}_{3\mathrm{qb}}$ into two parts, where $A$ contains the eigenstates of $\tilde{H}_{0}$, being $\{\ket{000}, \ket{111}\}$ and a segment $B$ containing all the remaining states of the computational basis.
We further reorganize the Hilbert space into these two blocks. A state in this new format is represented as
\begin{equation}
\label{eq:ketblock}
\ket{\psi}=\begin{pmatrix}
                       C_A \\
                    C_B
                      \end{pmatrix},
\end{equation}

where $C_A$ is a $2\times 1$ vector of $\ket{\psi}$ with components in the $A$ subspace, and $C_B$ is a $6\times 1$ vector related with the $B$ subspace. Operators also has an associated block representation in this format. For an operator $O$ given by
\begin{equation}
 \hat{O}  =  \begin{pmatrix}
          \hat{O}^{AA} & \hat{O}^{AB} \\
          \hat{O}^{BA} & \hat{O}^{BB}
         \end{pmatrix}.
\end{equation}
In this case, $\hat{O}^{AA}$ is a $2\times 2$ matrix in the $A$ subspace and $\hat{O}^{BB}$ is a $6\times 6$ matrix. The coupling between $A$ and $B$ is given by $\hat{O}^{AB}$, a $2\times 6$ operator, and $\hat{O}^{BA}$, a $6\times 2$ matrix. Following this definition, we consider $\tilde{\hat{H}}=\hat{O}$ acting over $\ket{\psi}$ in Eq.~\ref{eq:ketblock}, obtaining the following system of coupled equations
\begin{eqnarray}
\label{sis1}
 \tilde{\hat{H}}^{AA}C_A + \tilde{\hat{H}}^{AB}C_B &=& E C_A\nonumber\\
 \tilde{\hat{H}}^{BA}C_A + \tilde{\hat{H}}^{BB}C_B &=& E C_B
\end{eqnarray}
Isolating $C_B$ from the second line above, and replacing in the first line, we obtain
\begin{equation}
\label{sis4}
 \{\tilde{\hat{H}}^{AA} + \tilde{\hat{H}}^{AB}(E - \tilde{\hat{H}}^{BB})^{-1} \tilde{\hat{H}}^{BA}\} C_A  =  E C_A.
\end{equation}
Note that Eq. (\ref{sis4}) describes the effective dynamics in subspace $A$ where
\begin{equation}
\label{defhamefeGHZtrue}
 \tilde{\hat{H}}_{\mathrm{eff}}^{\mathrm{GHZ}}  =  \tilde{\hat{H}}^{AA} + \tilde{\hat{H}}^{AB}(E - \tilde{\hat{H}}^{BB})^{-1} \tilde{\hat{H}}^{BA},
\end{equation}
is the effective Hamiltonian operator.

At this point, we define
\begin{equation}\label{defOmega}
 \Omega_{\mathrm{GHZ}}  = \bra{111 } \tilde{\hat{H}}_{\mathrm{eff}}^{\mathrm{GHZ}} \ket{000},
\end{equation}
which means the problem of calculate the effective coupling behind the formation of the GHZ states becomes the problem of computing the expression above, considering $\Delta\ll J$.

To begin the calculation, we start with finding the explicit form of the operators $\tilde{\hat{H}}^{AA}$, $\tilde{\hat{H}}^{AB}$, $\tilde{\hat{H}}^{BA}$, $\tilde{\hat{H}}^{BB}$. By defining the projector operators $P$ and $Q$, such that
\begin{eqnarray}
\label{defPQ}
 \hat{P}&=&\ket{000}\bra{000} + \ket{111} \bra{111},\nonumber\\
 \hat{Q}&=&I - \hat{P},
\end{eqnarray}
where $I$ stands for the identity operator. Using the definitions for $\hat{H}_0$ and $\hat{V}$, Eq.~(\ref{defH0}) and Eq.~(\ref{perturbation}), we obtain
\begin{eqnarray}
\label{defHAsBs}
 \tilde{\hat{H}}^{AA}&=&\hat{P}(\tilde{\hat{H}}_0 + \hat{V})\hat{P},\nonumber\\
 \tilde{\hat{H}}^{BB}&=&\hat{Q}(\tilde{\hat{H}}_0 + \hat{V})\hat{Q},\nonumber\\
 \tilde{\hat{H}}^{AB}&=&\hat{P}(\tilde{\hat{H}}_0 + \hat{V})\hat{Q},\nonumber\\
 \tilde{\hat{H}}^{BA}&=&\hat{Q}(\tilde{\hat{H}}_0 + \hat{V})\hat{P}.
\end{eqnarray}
Here the operators $\tilde{\hat{H}}^{AA}$ and $ \tilde{\hat{H}}^{BB}$ are diagonal matrizes, while $\tilde{\hat{H}}^{AB}$ and $ \tilde{\hat{H}}^{BA}$ are nondiagonal, depending only of the parameter $\Delta$. Substituting Eq. (\ref{defhamefeGHZtrue}) into Eq. (\ref{defOmega}) leave us with
\begin{equation}
\label{defOmega2}
\Omega_{\mathrm{GHZ}}=\bra{111}\tilde{\hat{H}}^{AB}(E - \tilde{\hat{H}}^{BB})^{-1} \tilde{\hat{H}}^{BA} \ket{000},
\end{equation}
where we used the fact that $\tilde{\hat{H}}^{AA}$ is diagonal in $A$. Using the second line in Eq.(\ref{defHAsBs}), we have
\begin{equation}\label{defUseful}
 \tilde{\hat{H}}^{BB}  =  \hat{Q}\tilde{\hat{H}}_0\hat{Q} + \hat{Q}\hat{V}\hat{Q}   \equiv   \tilde{\hat{H}}_0^{BB} + \hat{V}^{BB}.
\end{equation}
Using Eq.~(\ref{defUseful}) we can write
\begin{equation}
 E - \tilde{\hat{H}}^{BB}  =  (E - \tilde{\hat{H}}_0^{BB})(I - (E - \tilde{\hat{H}}_0^{BB})^{-1}\hat{V}^{BB}),
\end{equation}
and thus it follows the identity
\begin{equation}\label{defInversa}
 (E - \tilde{\hat{H}}^{BB})^{-1}  = \{I - (E - \tilde{\hat{H}}_0^{BB})^{-1}\hat{V}^{BB}\}^{-1}(E - \tilde{\hat{H}}_0^{BB})^{-1}.
\end{equation}
Substituting (\ref{defInversa}) into eq. (\ref{defOmega2}), we have
\begin{eqnarray}
\label{defOmega3}
 \Omega_{\mathrm{GHZ}}&=& \bra{111} \tilde{\hat{H}}^{AB} \{I - (E - \tilde{\hat{H}}_0^{BB})^{-1}\hat{V}^{BB}\}^{-1} \nonumber\\
 &&\times(E - \tilde{\hat{H}}_0^{BB})^{-1} \tilde{\hat{H}}^{BA} \ket{000}.
\end{eqnarray}
At this point, we substitute the operator $\{I - (E - \tilde{\hat{H}}_0^{BB})^{-1} V^{BB}\}^{-1}$ for a Taylor expansion written as
\[
\{I - (E - \tilde{\hat{H}}_0^{BB})^{-1} \hat{V}^{BB}\}^{-1} = I + (E - \tilde{\hat{H}}_0^{BB})^{-1} \hat{V}^{BB},
\]
which is valid for $\Delta$ small if compared with the Coulomb coupling $J$. By substituting this term in the definition of $\Omega_{\mathrm{GHZ}}$ and inserting two identity operators we obtain Eq~(\ref{defOmega5}) from the main text. Notice that one can follow the same procedure to determine $\Omega_{\mathrm{FLIP}}$ by changing the segment $A$ for the subspace with $E=-2J$, with elements $\{\ket{101}$, and $\ket{101}\}$.

\end{document}